\documentclass[a4paper,11pt]{article}
\usepackage{pos}

\usepackage{url}

\usepackage{bm}% bold math
\newcommand{\qT}{q_T}

\title{Neural-Network extraction of TMDs with SIDIS data}

\author*[a]{Matteo Cerutti}

\affiliation[a]{Université Paris-Saclay - CEA - IRFU,\\
  91191 Gif-sur-Yvette, France}
  
\emailAdd{matteo.cerutti@cea.fr}

\abstract{
A first global analysis of unpolarized Transverse-Momentum-Dependent (TMD) distributions based on a neural-network (NN) parametrization is presented. Drell-Yan (DY) and semi-inclusive deep inelastic scattering (SIDIS) data are simultaneously included at next-to-next-to-next-to-leading logarithmic (N$^3$LL) accuracy.
The results indicate that the inclusion of SIDIS data leads to broader unpolarized TMD PDFs compared to a DY-only NN extraction. The associated uncertainties are reduced with respect to the DY-only case, while remaining larger than the ones obtained using traditional models.
These results demonstrate the potential of flexible NN parametrizations in reducing model dependence and provide guidance for future high-precision measurements at Jefferson Lab and the Electron-Ion Collider.
}

\FullConference{The 33rd International Workshop on Deep Inelastic Scattering and Related Subjects (DIS2026)\\
4 - 8 May 2026\\
Bologna, Italy\\}

\begin{document}
\maketitle

\section{Introduction}

Transverse-Momentum-Dependent (TMD) distributions provide a three-dimensional map of hadron structure in momentum space. Drell-Yan (DY) lepton-pair production in hadron-hadron collisions provides information on quark TMD parton distribution functions (TMD PDFs)~\cite{Arnold:2008kf}, while semi-inclusive Deep Inelastic Scattering (SIDIS) measurements provide simultaneous access to TMD PDFs and fragmentation functions (TMD FFs)~\cite{Bacchetta:2006tn}. Gluon TMDs can be accessed through other processes, like Higgs and/or quarkonia production~\cite{Anedda:2026cox}.

Despite significant progress in improving the perturbative accuracy and increasing the global dataset, there are tensions between current extractions of unpolarized quark TMD PDFs from different global fits~\cite{Moos:2025sal,Bacchetta:2024qre,Bacchetta:2022awv}. This is in part due to the limited flexibility of standard parametrizations and motivates the exploration of more flexible approaches in modeling the nonperturbative part of TMDs. A proof-of-concept study on DY data has recently demonstrated the feasibility of applying Neural Networks (NNs) to reduce this parameterization bias~\cite{Bacchetta:2025ara}.

In this contribution, I present an extension of Ref.~\cite{Bacchetta:2025ara} to a global TMD analysis. I focus in particular on the impact of SIDIS measurements, assessing their potential to constrain TMD distributions in a global framework. These results are particularly relevant for maximizing the physics potential of forthcoming experimental programs at future facilities, such as SoLID~\cite{Cerutti:2025xrp} at Jefferson Lab and the Electron-Ion Collider~\cite{AbdulKhalek:2021gbh}.

\section{Framework}

In the DY process $h_A + h_B \rightarrow \ell^+ + \ell^- + X$ two
hadrons with mass $M$ and four-momenta $P_A$ and $P_B$ collide with
center-of-mass energy squared $s=(P_A+P_B)^2$, and produce a lepton
pair with four-momentum $q$ and invariant mass $Q$.
If $Q \gg M$ and the transverse
momentum $\boldsymbol{q}_T$ relative to the collision axis is such that $|\boldsymbol{q}_T| \equiv q_T \ll Q$, the differential cross section reads
\begin{equation}
  \begin{split}
\label{e:DYZ_xsec}
&\frac{d\sigma^{\text{DY}}}{d q_T\, dy\, dQ}  = \frac{8 \pi \alpha^2 q_T}{9 Q^3}\, {\cal P}\, x_A\, x_B \, {\cal H}^{\text{DY}}(Q,\mu)\, \sum_a c_a(Q^2)
  \\
&\times \int_0^{\infty}\!\! db_T \, b_T J_0\big( b_T q_T \big) \hat{f}_1^a (x_A, b_T^2; \mu, \zeta_A) \, \hat{f}_1^{\bar{a}} (x_B,  b_T^2; \mu, \zeta_B),
  \end{split}
\end{equation}
where
$\alpha$ is the electromagnetic coupling,
$y=\ln\sqrt{(q_{0}+q_{z})/(q_{0}-q_{z})}$ is the lepton-pair rapidity,
${\cal P}$ is the phase-space-reduction factor due to possible fiducial cuts on the final-state leptons.
$x_{A,B} = Q e^{\pm y}/\sqrt{s}$ are
the longitudinal momentum fractions of the quarks in the initial state,
${\cal H}^{\text{DY}}$ is a perturbative hard factor, and the sum runs over all active quark flavors $a$ with $c_{a}$ the quark electroweak charges.

In Eq.~\eqref{e:DYZ_xsec}, $\hat{f}^a_{1}$ denotes the Fourier transform of the
unpolarized TMD PDF of quark flavor $a$. It depends on the longitudinal
momentum fraction $x$ and on $b_T=|\bm{b}_T|$,
Fourier conjugate to the quark intrinsic transverse momentum $\bm{k}_\perp$. Also, it depends on the UV renormalization scale $\mu$ and on the rapidity scale $\zeta$ (with $\zeta_A \zeta_B = Q^4$)~\cite{Collins:2011zzd}.

We also consider the SIDIS process  $\ell + h \rightarrow \ell' + h' + X$ where a lepton and a hadron with mass $M$ and four-momenta $l$ and $P$ collide, respectively, and inclusively produce a hadron $h'$ via the exchange of a vector boson with total four-momentum $q$ and invariant mass $q^2=-Q^2$. When $Q^2\gg M^2$ and $q_T \ll Q$, the differential cross section can be written as
\begin{equation}
  \begin{split}
\label{e:SIDIS_xsec}
&\frac{d\sigma^{\text{SIDIS}}}{d q_T\, dx\, dz\, dQ}  =\frac{16\pi^2\, \alpha^2\, z^2\, \qT}{x\, Q^3} \, \bigg (1 + \bigg( 1 - \frac{Q^2}{xs}  \bigg)^2 \bigg ) \,  \frac{x}{2\pi}\, {\cal H}^{\text{SIDIS}}(Q,\mu) \sum_a e_a^2
  \\
&\times \int_0^{\infty}\!\! db_T \, b_T J_0\big( b_T q_T \big) \hat{f}_1^{a/h} (x, b_T^2; \mu, \zeta_A) \, \hat{D}_1^{a \rightarrow h'} (z,  b_T^2; \mu, \zeta_B),
  \end{split}
\end{equation}
where $x$ and $z$ are the longitudinal momentum fractions in the incoming parton (w.r.t the parent hadron) and the outgoing hadron (w.r.t. the struck quark), respectively, and $\hat{D}_1^{a \rightarrow h'}$ is the TMD FF of a parton $a$ fragmenting into a hadron $h'$.

The structure of a TMD PDF is the following:
\begin{equation}
\hat{f}_1^{a/h} (x, b_T^2; \mu, \zeta) = [ C \otimes f_1 ]^a(x, b_*(b_T), \mu_{b_*}) \exp \bigg \{ \int_{\mu_{b_*}}^\mu \frac{d \mu'}{\mu'} \gamma(\mu', \zeta)\bigg \} \bigg (\frac{\zeta}{\mu^2_{b_*}} \bigg )^{K(\mu_{b_*})} \, f_{\rm NP}(x, b_T,\zeta) \,
\end{equation}
where $C$ is the perturbatively calculable coefficient function providing the matching onto collinear PDFs $f_1$ at small $b_T \ll 1 / \Lambda_{QCD}$, $\mu_b = 2 e^{-\gamma_E} / b_T$ is the natural choice for the TMD initial scale, $\gamma$ and $K$ are the anomalous dimensions of the evolution in $\mu$ and $\zeta$, respectively.
In order to avoid hitting the Landau pole when $b_T$ becomes large, the variable $b_T$ is replaced with $b_*(b_T)$, known as ``$b_*$ prescription'', which provides the transition between perturbative and nonperturbative physics~\cite{Collins:1984kg,Cerutti:2026apy}. The information on the genuine large-$b_T$ region is then encoded in a nonperturbative function $f_{\rm NP}$, which parametrizes long-distance QCD dynamics.

Perturbative ingredients are included consistently up to next-to-next-to-next-to-leading logarithmic accuracy (N$^3$LL). Collinear PDFs and FFs are provided through the LHAPDF library, using MSHT and DSS sets, respectively. Their uncertainties are incorporated consistently. SIDIS predictions include normalization factors computed prior to the fit according to Ref.~\cite{Bacchetta:2022awv}.

Nonperturbative components of TMD PDFs and FFs are parametrized as 
\begin{equation}
\begin{split}
& f_{ NP}(x, b_T; \zeta) = \frac{\mathbb{NN}(x, b_T)}{\mathbb{NN}(x,0)} \exp\left[ -g_2^2 b_T^2 \log \left(\frac{\zeta}{Q_0^2} \right) \right] \,, \\
& D_{ NP}(z, b_T; \zeta) = \frac{\mathbb{NN}(z, b_T)}{\mathbb{NN}(z,0)} \exp\left[ -g_2^2 b_T^2 \log \left(\frac{\zeta}{Q_0^2} \right) \right] \,,
\end{split}
\label{e:NNmodel}
\end{equation}
where $f_{\rm NP}$ is split into an ``intrinsic''
nonperturbative part, entirely parametrized by a NN, and the nonperturbative contribution to evolution, encoded in the exponential function (with $Q_0 = 1$~GeV). The two different NNs have architecture
$[2,10,1]$, \textit{i.e.} with two inputs corresponding to $x$ (or $z$ for $D_{NP}$) and $b_T$, $10$ hidden nodes, and one output node. The activation function is a sigmoid.
Note that $f_{ NP} \to 1$ for $b_T \to 0$ and $f_{\rm NP} \ll 1$ for large $b_T$. In this setup, the number of free parameters is 83, 41 for $f_{\rm NP}$, 41 for $D_{NP}$, and one ($g_2$) for the evolution.
To mitigate the risk of overfitting, the dataset is divided into training and validation samples of the same size.

\section{Results}

We discuss here the results of the simultaneous extraction of TMD PDFs and FFs obtained with the NangaParbat fitting framework~\cite{NangaParbat}. The global dataset includes low-invariant-mass DY measurements from Fermilab, high-energy $Z$-boson $q_T$-differential cross sections from RHIC and LHC, and SIDIS multiplicities from HERMES and COMPASS collaborations. Kinematic cuts are imposed to fulfill the requirements of TMD factorization. For DY data, the analysis is restricted to data at $q_T/Q < 0.2$, while the cut on SIDIS data is more involved (see Refs.~\cite{Bacchetta:2024qre,Bacchetta:2022awv}). The total number of data points in the fit is 2029 (482 for DY, 1547 for SIDIS).

\begin{figure}[h]
    \centering
    \includegraphics[width=0.495\textwidth]{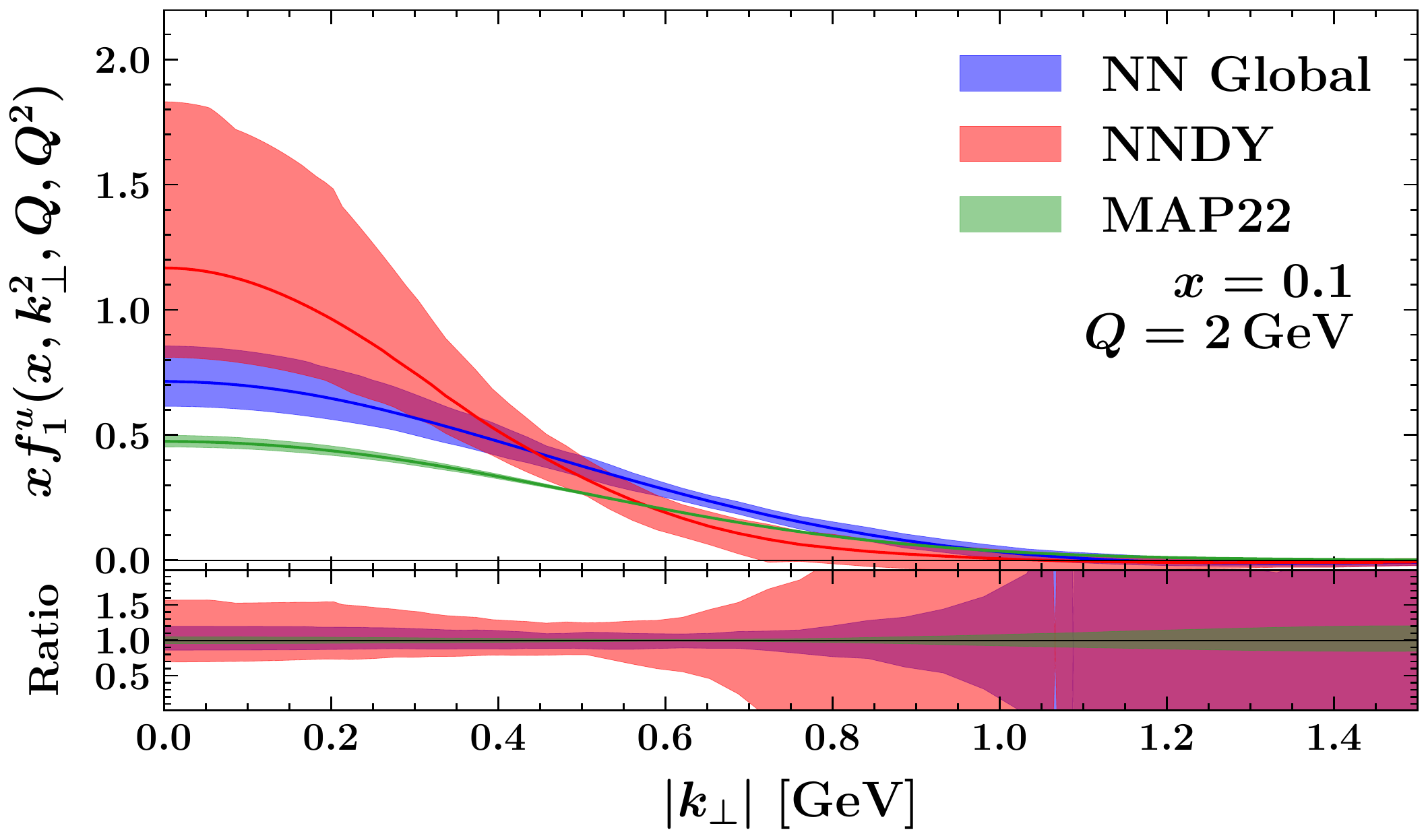}
    \includegraphics[width=0.495\textwidth]{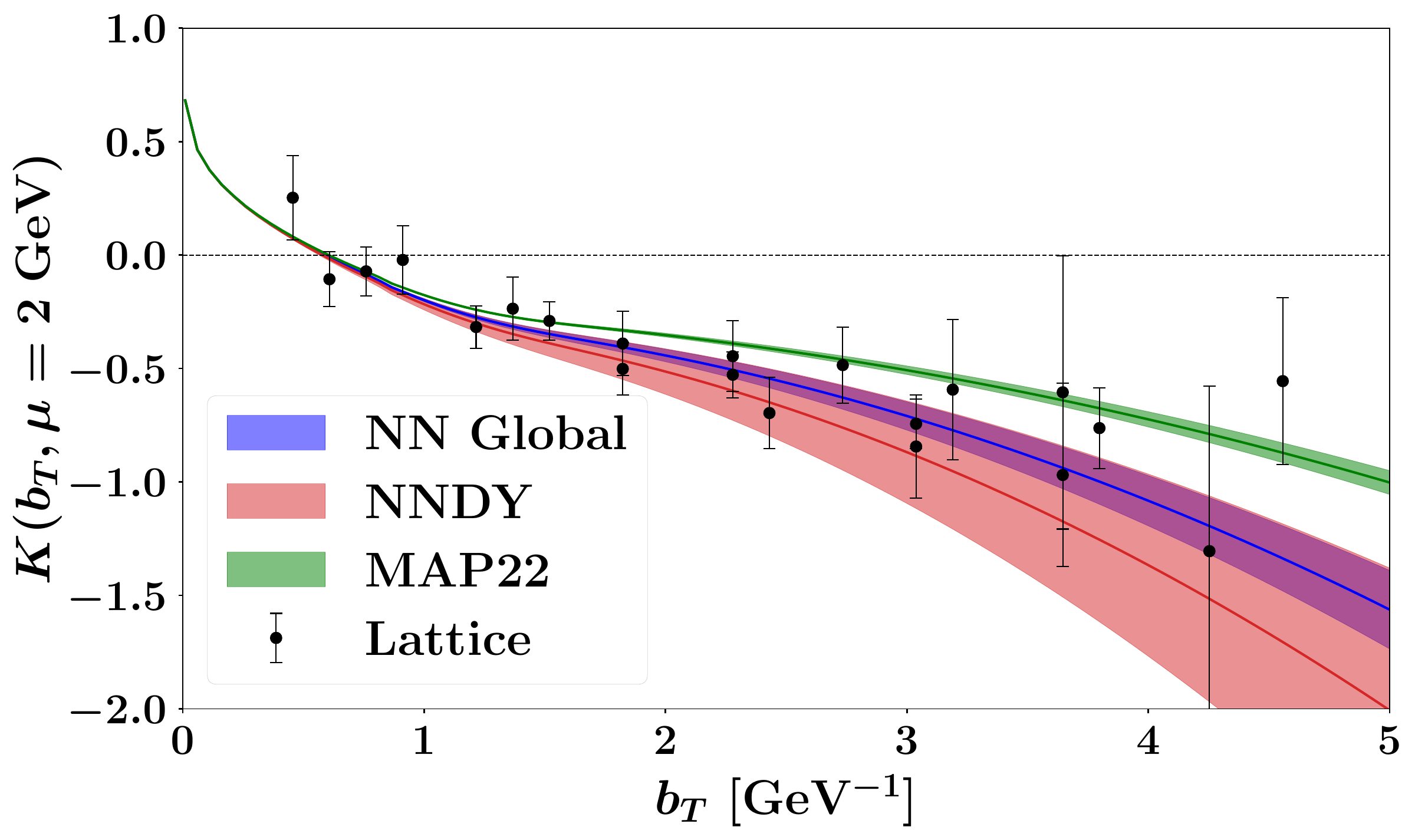}
    \caption{Comparison of the unpolarized $u$-quark TMD PDF at $x=0.1$ and $\mu=\sqrt{\zeta} = 2$ GeV (left panel) and the Collins-Soper kernel (right panel) obtained in this work (blue band) with NNDY (red band) and MAP22 extractions (green band). Uncertainty bands represent one-$\sigma$ uncertainties.}
    \label{f:TMDs}
\end{figure}

The propagation of experimental errors in the fit is performed via the bootstrap method, namely by fitting 100 Monte Carlo replicas of the experimental data. 
The values of the best-fit parameters are obtained by minimizing a $\chi^2$ function.
The quality of the fit is represented by the $\chi^2$ value normalized by the number of data points $N_{\rm dat}$. The resulting value is $\chi^2 / N_{\rm dat} = 1.01$, indicating that the analysis is able to simultaneously describe DY and SIDIS datasets.
In Fig.~\ref{f:TMDs}, we compare the extracted quantities with previous analyses. In the left panel, we compare the results of this NN fit (blue band), the NNDY~\cite{Bacchetta:2025ara} fit (red band) and the MAP22~\cite{Bacchetta:2022awv} fit (green band) for the unpolarized $u$-quark TMD PDF at $x=0.1$ and $\mu=\sqrt{\zeta} = 2$ GeV as a function of the intrinsic transverse momentum $|\bm{k}_\perp|$. In the right panel, the comparison is for the anomalous dimension of the rapidity evolution, known as Collins-Soper (CS) kernel, as a function of $b_T$ along with lattice calculations used in Ref~\cite{Avkhadiev:2025wps}.
Uncertainty bands represent 68\% uncertainties.
In the left panel, we observe that the TMD PDF from this analysis is broader than the one from the NNDY extraction. This broadening effect is likely due to the inclusion of SIDIS data in the fit.
This effect is also observed in the MAP22 extraction with traditional parametrizations.
TMD PDFs from this analysis have smaller uncertainties compared to those from the NNDY extraction, but larger than those from MAP22, indicating that the increased flexibility of the NN model allows for a more robust assessment of uncertainties.
In the right panel, we observe that the CS kernel from the NN Global analysis is in good agreement with the NNDY extraction and with the shown lattice calculations. In particular, at large $b_T$ the result from this analysis is smaller than that from the NNDY extraction, which is consistent with the broader TMD PDFs observed in the left panel. This result indicates that there are no significant tensions between DY and SIDIS data in the extraction of the nonperturbative TMD evolution.

\clearpage

\section*{Acknowledgments}
The author would like to thank the organizers for creating a stimulating environment throughout the conference and Valerio Bertone for his encouragement to complete these proceedings, as well as for his valuable comments and suggestions during their preparation.

\end{document}